\documentclass[preprint,showpacs,preprintnumbers,amsmath,amssymb]{revtex4}
\usepackage{graphicx}% Include figure files
\usepackage{dcolumn}% Align table columns on decimal point
\usepackage{bm}% bold math
\begin{document}

\title{Stretching DNA in hard-wall potential channels}

\author{Marco Zoli}

\affiliation{School of Science and Technology \\  University of Camerino, I-62032 Camerino, Italy \\ marco.zoli@unicam.it}

\date{\today}

\begin{abstract}
A three dimensional mesoscopic model is applied to study the properties of short DNA chains in a confining environment. The cylindrical channel is represented by a hard-wall repulsive potential incorporated in the system Hamiltonian. The macroscopic helical parameters are computed performing statistical averages over the ensemble of microscopic base pair fluctuations. The average molecule elongation, measured by the end-to-end distance, is derived as a function of the channel potential parameters both for a homogeneous and a heterogeneous chain. The overall results suggest that the mesoscopic model, with the channel potential term, yields consistent quantitative estimates for the stretching and twisting of short chains.
\end{abstract}


\maketitle

The predictability of the Watson-Crick base pairing is crucial for programmable connections of DNA fragments leading to self-assembly of 2D and 3D structures in devices and functional materials with biomedical applications \cite{seeman17,yang16}. 
The formation of these complex constructs highlights the intrinsic flexibility of the DNA molecule whose axis can be stretched at scales involving a few base pairs, hence shorter than its standard persistence length, i.e. $50$ nm \cite{io18b}. The DNA stretching properties have been extensively analyzed over the last decades by single molecule micro-manipulation techniques sampling the DNA extension as a response to applied loads \cite{busta06} or to proteins which deform the helical structure upon binding \cite{biton18}.  

The correlation between helical shape and stretching is also manifest when DNA is constrained in pores whose size shrinks the space available to the base pair fluctuations.  DNA nano-channels confinement in fluidic chips is emerging as a useful tool for accurate single molecule sequencing and genome mapping. To get high resolution DNA analysis one expects the molecule to be fully stretched and, in principle, this is achieved once the channel width is slightly broader than the helix diameter i.e. $\sim 20$\AA. On the other hand the fabrication of long nanofluidic chips is technically difficult already for widths of order $10$ nm and even more so for smaller sizes \cite{xia08}. 
The physics of long DNA molecules in nano-channels is generally described in the framework of polymer theory \cite{gennes77,odijk83} and bead-spring models \cite{chen13,pablo07}. However more detailed models accounting for the helical structure are required at those short length scales in which the relation between molecule stretching and shape is governed by the forces acting at the level of the single base pairs. 

In this regard mesoscopic Hamiltonian models may provide a better understanding of the physical properties of DNA once the molecule is placed in an environment which spatially reduces the base pair fluctuations. Accordingly we propose here a statistical mechanical approach  for the confined DNA chain which incorporates the bending and twisting fluctuations between adjacent base pairs and therefore describes the molecule helicity. The novelty of this study lies in the analysis of the effects brought about by a model potential simulating a single molecule driven through a cylindrical pore which uniformly constrains the amplitude of the base pair radial fluctuations.
As a consequence the molecule stretches and its elongation, measured by the average end-to-end distance, is computed as a function of the potential parameters. In line with the above mentioned single molecule experiments, our computational scheme admits that the DNA chain may change its torsional conformation upon stretching and establishes a  correlation between molecule size and helical shape, the latter being measured by the average number of base pairs per helix turn.
The results, presented for short homogeneous and heterogeneous chains, corroborate the choice of the channel potential for quantitative estimates of the molecule properties in narrow channels.

\section{ Linear DNA in a channel}

\begin{figure}
\includegraphics[height=7.0cm,width=8.0cm,angle=-90]{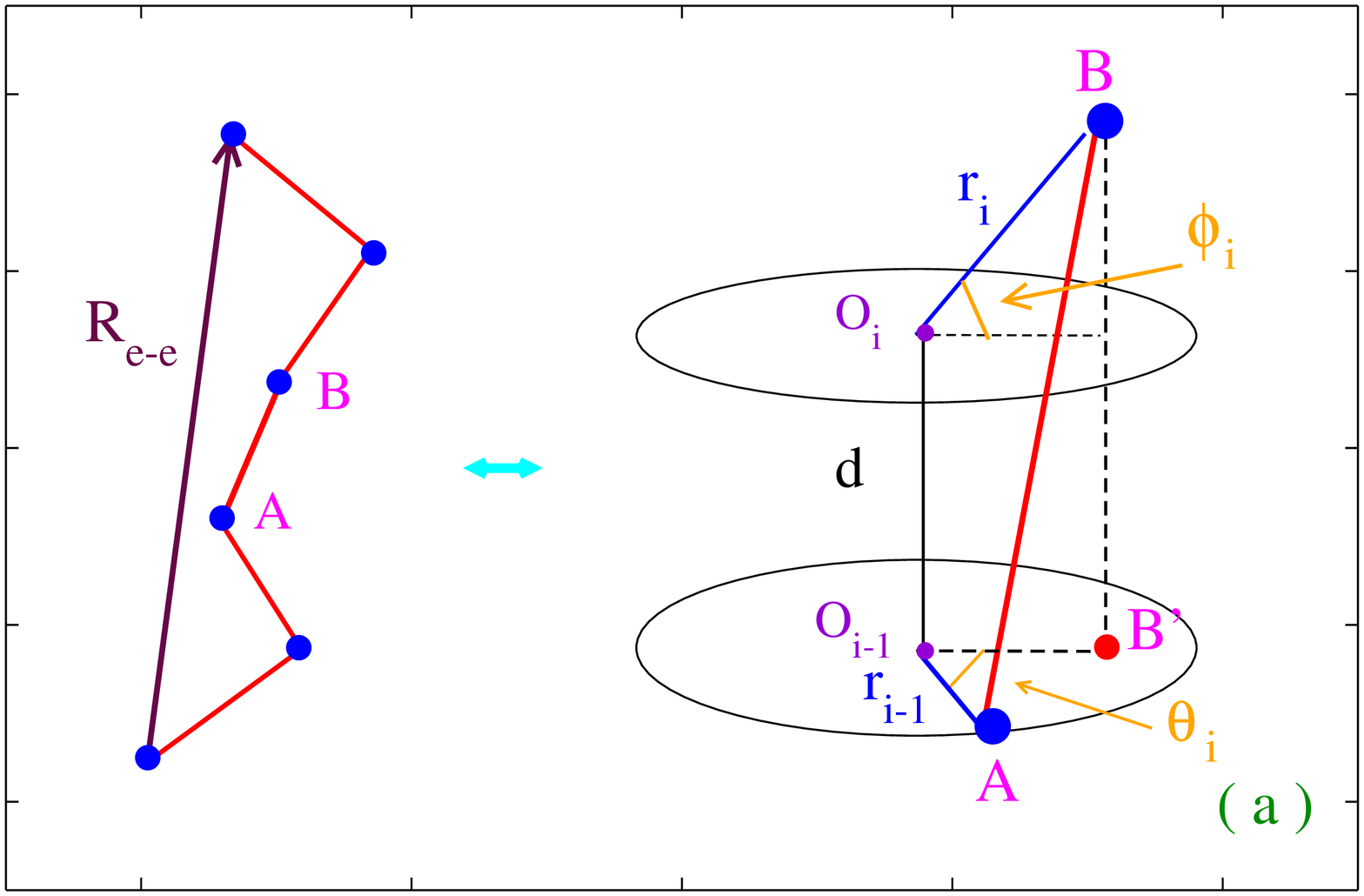}
\includegraphics[height=7.0cm,width=8.0cm,angle=-90]{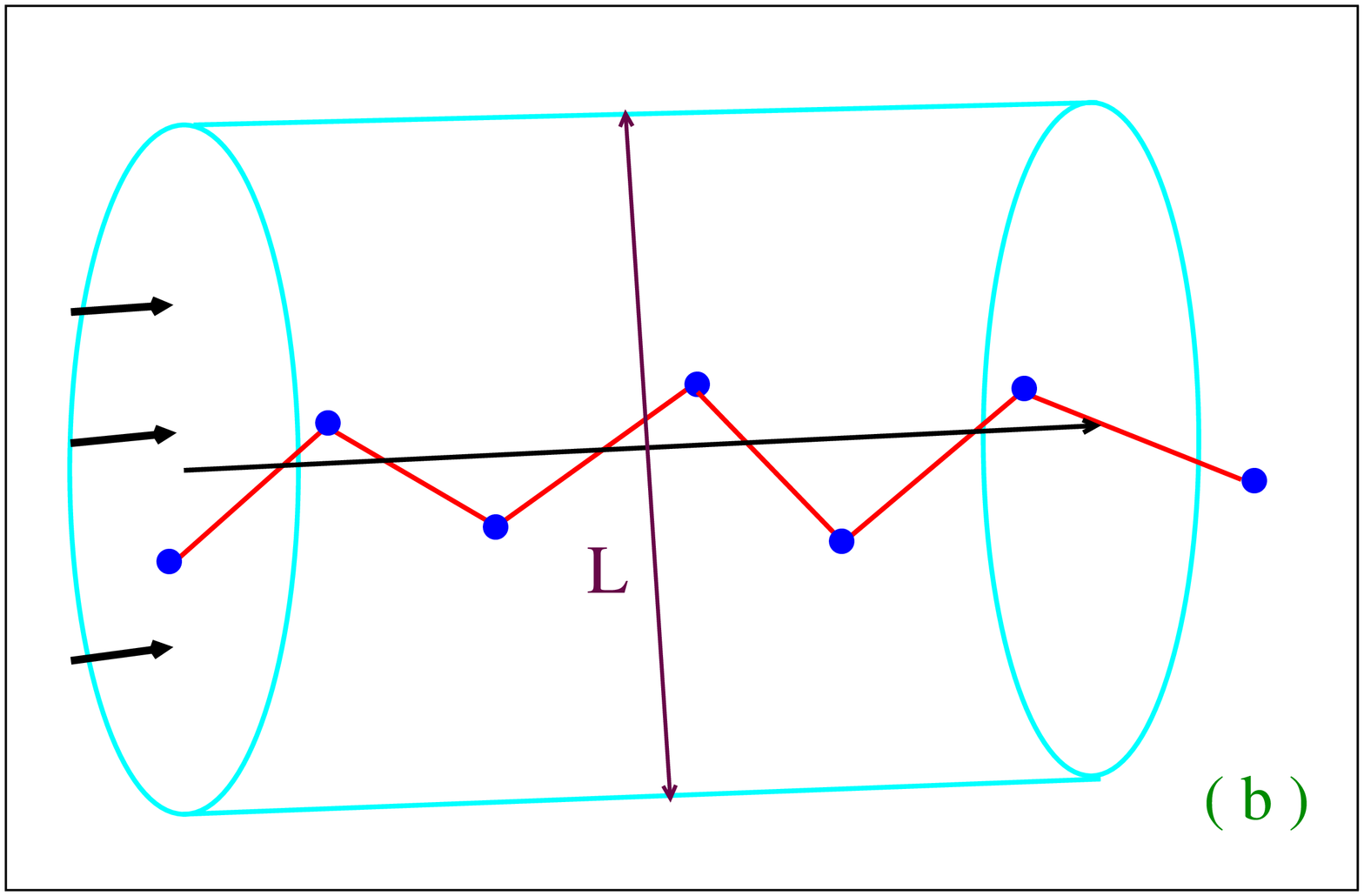}
\caption{\label{fig:1}(Color online)  
(a) Schematic of the model for a helical chain with $N$ point-like base pairs. $r_{i}$ is the inter-strand distance between the mates of the $i-th$ base pair. It is measured with respect to the point $O_i$ lying along the helix mid-axis. The $O_i$'s are separated by a constant $d$. The angles $\theta_i$ and $\phi_{i}$ define the local twist and bending between neighboring base pairs. $\overline{AB}$ is the distance between the radial displacements $r_{i}$, $r_{i-1}$. 
$R_{e-e}$ is the end-to-end distance. (b) The helical chain flows through a cylindrical channel of tunable diameter $L$ which uniformly confines the base pair fluctuations. By reducing $L$, the molecule stretches due to a stronger confinement.
}
\end{figure}

We begin with a geometrical model for the open ends helical chain with $N$ base pairs as shown in Fig.~\ref{fig:1}(a).
The blue dots denote the tips of the radial fluctuation vectors $\,r_i$'s measured with respect to the points $O_i$'s lying along the helix mid-axis.
By suppressing the radial fluctuations between  complementary pair mates, one would recover a freely jointed model of beads arranged along a linear chain with constant rise distance $d$.
As adjacent base pairs are twisted and bent by the variables $\theta_i$ and $\phi_i$ respectively, the three dimensional model accounts for fluctuational effects also on the angular degrees of freedom. For vanishing bending fluctuations, the present model maps onto a fixed-plane twisted representation (i.e. the ovals in Fig.~\ref{fig:1}(a)) previously applied to studies of bubble formation and thermodynamics of short chains \cite{io11}.

Note that: \textit{i)}  each blue dot represents precisely a nucleotide, i.e, the monomer unit comprising a base pair and the attached sugar-phosphate group on the molecule backbone. A description of the internal structure of base pairs and nucleotides, which would require full atomistic models, is not essential to our purposes and beyond the scope of coarse-grained analysis; \textit{ii)} the distance $\overline{d_{i,i-1}}$ between adjacent base pairs ($\overline{AB}$   in Fig.~\ref{fig:1}(a)) in general varies along the stack and depends on the local radial and angular fluctuations.  Then our model is an effective 3D helical representation of the double stranded molecule whose stability is governed by the intra-strand and inter-strand forces at the level of the base pairs as described below. The global size of the molecule, central to our investigation, is measured by the end-to-end distance $R_{e-e}$. 
As depicted in Fig.~\ref{fig:1}(b), the 3D helical chain is driven through a channel, e.g., by a hydrodynamic flow or electric field,  which shrinks the radial base pair fluctuations and also constrains the angular variables thus affecting the overall shape of the molecule. Here we assume that the channel is a cylinder with diameter $L$. As the mid-axis of the helical molecule, trapped in the pore, coincides with the cylinder axis, the $\,r_i$'s are uniformly compressed by the environment.

\section{Mesoscopic Hamiltonian}

The model in Fig.~\ref{fig:1} is studied by a Hamiltonian $H$ which is the sum of two contributions: a mesoscopic Hamiltonian for the ds-DNA molecule ($H_{mol}$) and a hard-wall potential due to the confining channel ($V_{ch}$).

Formally the DNA chain is modeled by:

\begin{eqnarray}
& &H_{mol} =\, H_a[r_1] + \sum_{i=2}^{N} H_b[r_i, r_{i-1}, \phi_i, \theta_i] \, , \nonumber
\\
& &H_a[r_1] =\, \frac{\mu}{2} \dot{r}_1^2 + V_{1}[r_1] \, , \nonumber
\\
& &H_b[r_i, r_{i-1}, \phi_i, \theta_i]= \,  \frac{\mu}{2} \dot{r}_i^2 + V_{1}[r_i] + V_{2}[ r_i, r_{i-1}, \phi_i, \theta_i]  \, . \nonumber
\\
\label{eq:01}
\end{eqnarray}

$\mu$ is the base pair reduced mass. 
As the first base pair along the stack has no preceding neighbour, the term $H_a[r_1]$ is taken out of the sum. 
$V_{1}[r_i]$ is the one particle potential modeling the hydrogen bonds between complementary pair mates while the intra-strand covalent forces between adjacent base pairs are described by the two particles potential $V_{2}[ r_i, r_{i-1}, \phi_i, \theta_i]$.  

Explicitly, $V_{1}$ and $V_{2}$ are written as:

\begin{eqnarray}
& &V_{1}[r_i]=\, V_{M}[r_i] + V_{Sol}[r_i] \, , \nonumber
\\
& &V_{M}[r_i]=\, D_i \bigl[\exp(-b_i (|r_i| - R_0)) - 1 \bigr]^2  \, , \nonumber
\\
& &V_{Sol}[r_i]=\, - D_i f_s \bigl(\tanh((|r_i| - R_0)/ l_s) - 1 \bigr) \, , \nonumber
\\
& &V_{2}[ r_i, r_{i-1}, \phi_i, \theta_i]=\, K_S \cdot \bigl(1 + G_{i, i-1}\bigr) \cdot \overline{d_{i,i-1}}^2  \, , \nonumber
\\
& &G_{i, i-1}= \, \rho_{i, i-1}\exp\bigl[-\alpha_{i, i-1}(|r_i| + |r_{i-1}| - 2R_0)\bigr]  \, . \nonumber
\\
\label{eq:02}
\end{eqnarray}

\textit{(1)} $V_{M}$ is the hydrogen bond Morse potential accounting for the inter-strand base pair interaction through its site dependent parameters, the pair dissociation energy $D_i$ and inverse length $b_i$. As $R_0 \sim 20 \AA$ is the bare helix diameter, the potential minimum corresponds to the absence of radial fluctuations, i.e., $|r_i| = R_0$. The hard core of $V_{M}$ accounts for the repulsive inter-strand electrostatic interactions due to negatively charged phosphate groups and provides a physical criterion to define the spatial range of the fluctuations contributing to the calculation of the partition function.
Precisely, for short radial fluctuations such as $\,|r_i| - R_0 < - \ln 2 / b_i$, the complementary pair mates would come too close to each other generating a large electrostatic energy, $V_{M}[r_i] >  D_i $. Hence, such fluctuations would have a small statistical weight in the partition function. Accordingly the code operates a truncation of the integration range excluding those fluctuations yielding a sizeable contraction of the helix diameter with respect to the bare value.
The one particle potential also contains a solvent contribution $V_{Sol}$ which has the effect to stabilize the base pair breaking associated to large radial fluctuations \cite{coll95}. In fact,  $V_{Sol}$ enhances by $D_i f_s$ the threshold for pair dissociation and, for large base pair fluctuations, introduces a hump (whose width is controlled by $l_s$) over the Morse plateau thus accounting for the strand recombination effects which may occur in solution. 
The parameter $f_s$ can empirically account for the salt concentration in the solvent which affects the thermodynamic parameters \cite{albu14} and contributes to shape the DNA conformation.  

\textit{(2)} $V_{2}[ r_i, r_{i-1}, \phi_i, \theta_i]$ is the stacking potential depending on the square of the distance $\overline{d_{i,i-1}}$ magnified in Fig.~\ref{fig:1}(a).
The harmonic force between adjacent base pairs is weighed by the elastic constant $K_S$ while the term $G_{i, i-1}$ accounts for the non linear stacking contributions tuned by the parameters $\rho_{i, i-1}$ and $\alpha_{i, i-1}$. Non linear intra-strand forces had been first introduced in a ladder Hamiltonian model (which lacks the angular variables) aimed to describe the sharpness of the thermally driven DNA denaturation \cite{pey93}. In that 1D model however the stacking potential displays a unphysical divergence whenever, because of thermal fluctuations, a hydrogen bond is broken and the corresponding base is unstacked thus reducing the overlap between adjacent $\pi$ electron clouds. In fact, under such circumstances, the distance $r_i - r_{i-1}$ becomes very large. This drawback is solved by our 3D model as the angular variables stabilize the double helix against thermal disruptions \cite{io12} thus ensuring a finite stacking energy physically associated to the stiffness of the sugar-phosphate covalent bonds. 

Furthermore, $\,V_{2}[ r_i, r_{i-1}, \phi_i, \theta_i]\,$ remains finite (non zero) also when all radial fluctuations are equal, i.e. for the translational mode. It follows that, in our model, the partition function does not diverge for equal and large $r_i$'s. This makes a substantial advantage with respect to ladder models which, instead, have to tackle the divergence of the partition function arising from the concomitant facts that \textit{i)} the two particle potential vanishes for equal and large fluctuations and \textit{ii)} the one particle potential is bounded for $r_i \rightarrow \infty $ \cite{zhang97,kalos09}.
 
For the homogeneous fragment studied hereafter we take the parameters, $D_i=\,60 meV$, $b_i= 3 \AA^{-1}$,  $f_s=\,0.1$, $l_s=\,0.5 \AA$, $K_S=\,10 mev \AA^{-2}$, $\rho_{i} \equiv \, \rho_{i, i-1} =\,1$, $\alpha_{i}\equiv \, \alpha_{i, i-1} =\, 2 \AA^{-1}\,$  used in previous studies and within the range of values consistent with available thermodynamic and elastic data  \cite{lucia98,campa98,krueg06,io18c}. In particular, the pair dissociation energy value is appropriate for $GC$ base pairs. Heterogeneous DNA chains can also be modeled via Eq.~(\ref{eq:02}) differentiating the parameter values for the specific $AT$ and $GC$ base pairs along the sequence \cite{io13,weber15} as done below. It is noted that the model potential does not discern a $AT$ from a $TA$ base pair.

The physical effect of the channel in Fig.~\ref{fig:1}(b) is introduced in the Hamiltonian through the potential:

\begin{eqnarray}
V_{ch}(r_i)  =\,
\left\{
\begin{matrix}
\gamma  \cdot \bigl| |r_i| - R_0 - \delta(L) \bigr|^{-1}  \hskip 0.6cm          |r_i| - R_0 < \delta(L)    \\ 
\infty                       \hskip 3.5cm          |r_i| - R_0 \geq  \delta(L) 
\end{matrix} 
\right .  
\label{eq:03}
\end{eqnarray}

where $\gamma$ and $\delta(L)$ are tunable parameters measuring the strength and the range of the force exerted by the cylinder walls on the base pairs motion. This choice for $V_{ch}(r_i)$ accounts for the fact that by narrowing the channel one truncates the phase space available to the base pair fluctuations thus reducing their statistical weight. $\gamma$ may also account for the electrostatic repulsion between DNA and channel if its walls are negatively charged. While $\delta(L)$ may not coincide with the diameter $L$, the relation $\, \delta(L) \propto L \,$ should hold.

Then, from Eqs.~(\ref{eq:01}),~(\ref{eq:03}), the total Hamiltonian of the system reads:

\begin{eqnarray}
H_T=\, H_a[r_1] + V_{ch}(r_1) + \sum_{i=2}^{N} \Bigl( H_b[..] + V_{ch}(r_i) \Bigr) \,. 
\label{eq:04}
\end{eqnarray}

with $H_b[..] \equiv H_b[r_i, r_{i-1}, \phi_i, \theta_i]$.

\section{Partition Function}

The model in Eq.~(\ref{eq:04}) is studied by a well established computational method largely described in previous works \cite{io14b,io16b} and briefly outlined here.
The method relies on the assumption that the fluctuational distances $r_i$  are temperature dependent paths. Hence, the system can be treated by the finite temperature path integral formalism \cite{fehi} where the path integral for a chain with $N$ base pairs is given by a sum over the paths $r_i(\tau)$ with the Euclidean time $\tau \in [0, \beta]$ and  $\beta=\,(k_B T)^{-1}$. $k_B$ is the Boltzmann constant and $T$ is the temperature. It follows that the statistical partition function $Z_N$ is obtained as an integral over paths obeying the closure condition, $\,r_i(0)=\,r_i(\beta )\,$, which can be enforced by the Fourier expansion, \, $r_i(\tau)=\, (r_0)_i + \sum_{m=1}^{\infty}\Bigl[(a_m)_i \cos( \frac{2 m \pi}{\beta} \tau ) + (b_m)_i \sin(\frac{2 m \pi}{\beta} \tau ) \Bigr] \,$. While the Fourier coefficients define in principle all possible choices of fluctuations for any base pair, the calculation includes in $Z_N$ a subset of $r_i$'s which fulfill the above described physical requirements.

Note that the path closure condition plays the role of the periodic boundary conditions (PBC) usually employed in transfer integral solutions of the Peyrard-Bishop DNA Hamiltonian \cite{zhang97,singh11}. In transfer integral methods however PBC are implemented either by introducing a fictitious base pair or by closing the open end chain into a loop. While both strategies may be suitable for long molecules, the application to short chains is questionable as the boundaries may significantly contribute to the properties of linear DNA. Instead, the path integral method does not incur this problem as the closure condition is imposed on the inverse temperature scale, not in the real space. Thus, the computation assumes an open ends molecule.

Then for the system in Eq.~(\ref{eq:04}), considering the integration over the radial and angular variables, $Z_N$ reads:

\begin{eqnarray}
& &Z_N=\, \oint Dr_{1} \exp \bigl[- A_a[r_1] \bigr]   \prod_{i=2}^{N}  \int_{- \phi_{M} }^{\phi_{M} } d \phi_i \int_{- \theta_{M} }^{\theta _{M} } d \theta_{i} \times  \,  \nonumber
\\
& &\oint Dr_{i}  \exp \bigl[- A_b [r_i, r_{i-1}, \phi_i, \theta_i] \bigr] \, , \nonumber
\\
& &A_a[r_1]= \,  \int_{0}^{\beta} d\tau \Bigl( H_a[r_1(\tau)] + V_{ch}[r_1(\tau)] \Bigr) \, , \nonumber
\\
& & A_b[r_i, r_{i-1}, \phi_i, \theta_i]= \,  \int_{0}^{\beta} d\tau \Bigl( H_b[..] + V_{ch}[r_i(\tau)] \Bigr) \, ,
\label{eq:05}
\end{eqnarray}

where $A_a[r_1]$ is the dimensionless action for the first base pair which is coupled only to the successive base pair via the action contribution $A_b[r_2, r_{1}, \phi_2, \theta_2]$.   The integration cutoffs on the bending and twisting variables are taken large enough to allow for local structural distortions \cite{kim14} which cause the overall flexibility of the molecule \cite{vafa} i.e., $\phi_{M}=\,\pi /6$ and $\theta_{M}=\,\pi /4$ respectively. In particular, as the bending fluctuations crucially affect the global size of the chain, the quantitative effect of $\phi_{M}$ will be discussed in the following.

$\oint {D}r_i$ is the measure of integration over the Fourier coefficients associated to the path $r_i(\tau)$:

\begin{eqnarray}
& &\oint {D}r_{i} \equiv {\frac{1}{\sqrt{2}\lambda_{cl}}} \int_{-\Lambda_{T}^{0}}^{\Lambda_{T}^{0}} d(r_0)_i \prod_{m=1}^{\infty}\Bigl( \frac{m \pi}{\lambda_{cl}} \Bigr)^2  \times \nonumber
\\
& &\int_{-\Lambda_{T}}^{\Lambda_{T}} d(a_m)_i \int_{-\Lambda_{T}}^{\Lambda_{T}} d(b_m)_i \, , \, 
\label{eq:06}
\end{eqnarray}

where  $\lambda_{cl}$ is the classical thermal wavelength.  Applying the normalization condition intrinsic to the path integral method \cite{io16b}: 

\begin{eqnarray}
\oint {D}r_i \exp\Bigl[- \int_0^\beta d\tau {\mu \over 2}\dot{r}_i(\tau)^2  \Bigr] = \,1 \, ,
\label{eq:07} \,
\end{eqnarray}

one derives the cutoffs on the Fourier coefficients integration: \, $\Lambda_{T}^{0}=\,\lambda_{cl}/\sqrt{2} $ and  $\Lambda_{T}=\,{{U \lambda_{cl}} / {m \pi^{3/2}}}$.  $U$ is a dimensionless parameter which controls the range of the temperature dependent radial fluctuations and can be set numerically by Eq.~(\ref{eq:07}). Thus, the bounds for the base pair fluctuational amplitudes are consistently determined in the path integration method and this makes a distinctive advantage over transfer integral and molecular dynamics simulations which necessarily operate somewhat arbitrary truncations of the phase space available to the base displacements \cite{zhang97}. 
Eq.~(\ref{eq:07}) also defines the free energy zero for a set of free base pairs and holds for any $\mu$. 
Consistently, the free energy does not depend on $\mu$.

Incidentally we note that by reducing $U$ in Eq.~(\ref{eq:06}) via a site dependent contraction factor, one could also simulate the effects of a crowder distribution which exerts a \textit{non-uniform} confinement on the base pairs in the chain. By this phenomenological approach, which does not require $V_{ch}(r_i)$ in Eq.~(\ref{eq:04}), one may account for the variable size of the crowders and study the changes in shape and size of the DNA molecule brought about by such distribution \cite{io19}.

Eqs.~(\ref{eq:05}),~(\ref{eq:06}) define the ensemble of base pair radial and angular fluctuations over which the statistical averages for the DNA macroscopic parameters are carried out. The computation is performed by increasing the number of integration paths in $Z_N$ up to get numerical convergence i.e, the configuration of thermodynamic equilibrium.

\section{Helical Parameters}

The DNA molecule is expected to modify the torsional conformation upon stretching whereby the latter may be caused by external forces, crowders or confining channels. To account for this correlation, we have devised a iterative computational scheme in which the twist variable $\theta_i$ ( Fig.~\ref{fig:1}(a) ) is measured from the ensemble averaged twist angle for the preceding base pair along the chain axis i.e., $<\theta_{i - 1}> $ with \, $< \theta_1 >\equiv \,0$. 

Formally: $\,\theta_i =\, <\theta_{i - 1}>  + 2\pi / h_j + \theta_{i}^{fl}$, \, where $\theta_{i}^{fl}$ is the twist fluctuation angle to be integrated in Eq.~(\ref{eq:05}) and $h_j$ is the variable number of base pairs per helix turn. Precisely $h_j$ is taken within a range ($j=\,1,...,J$) around the experimental value, $h^{exp}=\,10.4$ for kilo base long DNA \cite{wang79} , thus admitting that the DNA short chain may change its helical conformation under the effect of the hard wall potential.
For a specific  $h_j$, we compute all the average twist angles in the chain and, from $< \theta_N >$, one derives the $j-th$ ensemble averaged helical repeat:
 
\begin{eqnarray}
< h >_{j}=\,\frac{2\pi N}{< \theta_N >} \, . 
\label{eq:08}
\end{eqnarray}

Performing the calculation for any $h_j$ in the range, we obtain a set ($< h >_{1},..., < h >_{J}$) of averaged helical repeats defining $J$ possible twist conformations among which the value ($< h >_{j^{*}}$) for the state of thermodynamic equilibrium is selected by minimization of the free energy $\, F=\, -\beta ^{-1} \ln Z_N$.
The accuracy of the technique depends on the mesh of sampled conformations. Taking $J=\,201$ with a fine partition step $\Delta h =\,0.0625$, we explore a broad range $h_j \in (6,\,18)$ of helical conformations clearly at the price of a high computational time \cite{io17}.
 
Eventually, for any twist conformation, we also compute the ensemble averaged end-to-end distance (Fig.~\ref{fig:1}(a))) measuring the molecule size as a function of the dimers stacking distances:

\begin{eqnarray} 
< R_{e-e} >_{j} =\, \biggl < \biggl| \sum_{i=2}^{N}  \overline{d_{i,i-1}} \biggr| \biggr >   \,.
\label{eq:09}
\end{eqnarray}

Likewise, $< R_{e-e} >_{j*}$ indicates the thermodynamic equilibrium value corresponding to the twist conformation $< h >_{j^{*}}$.

Note that, by virtue of the path integration method incorporating temperature dependent radial fluctuations, the thermal effects are transferred also to the macroscopic average parameters in Eqs.~(\ref{eq:08}),~(\ref{eq:09}). Thus, the method would permit to monitor e.g. the thermally driven helix unwinding and the relation between helical twist and stretching as a function of temperature. All calculations here presented are carried out at room temperature.

\section{ Results}

\begin{figure}
\includegraphics[height=7.0cm,width=8.0cm,angle=-90]{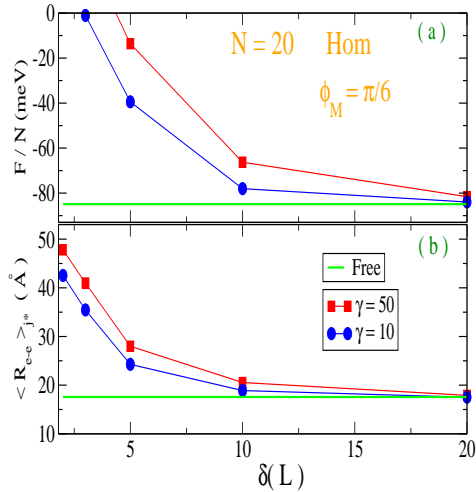}
\caption{\label{fig:2}(Color online)  (a) Free energy per base pair and (b) average end-to-end distance, for a homogeneous chain ($20$ base pairs) in a confining channel.  The effect of the channel diameter is tuned by $\delta(L)$ (units $\AA$). $\gamma$ (units $meV \AA^{-1}$) measures the strength of the hard wall potential. The free molecule values are also reported. Both the free energy and the end-to-end distance are calculated for the equilibrium twist conformation.
}
\end{figure}

As a first application of the method, a homogeneous fragment of ($N=\,20$) $GC$ base pairs under confinement is considered in Fig.~\ref{fig:2}. The free energy per base pair and the equilibrium average end-to-end distance are plotted as a function of  $\delta(L)$ for two choices of the interaction strength $\gamma$. The  $F/N$ values shown in Fig.~\ref{fig:2}(a) are on the negative axis with the free energy zero being defined via Eq.~(\ref{eq:07}). 
The results for the free molecule ($V_{ch} \, \equiv 0$) are given for comparison (green line).  Our analysis is purely predictive as experimental data for the quantitative stretching of confined short chains are currently not available.

While for $\delta(L) \sim 20\AA$ the obtained values for the confined and free molecule essentially overlap, a significant free energy increase is found for $\delta(L) < 10\AA \,$ clearly more pronounced for the model with larger $\gamma$.  Consistently, the entropic reduction driven by a narrower pore is accompanied by a substantial stretching of the chain as shown in Fig.~\ref{fig:2}(b). While the largest calculated stretching corresponds to $\delta(L)=\, 2 \AA$, even stronger confinements can be achieved by further tweaking the channel parameters. It is reminded that, the length of a straight chain, with $N=\,20$, is $\sim 65 \AA$. 

One may wonder to which extent the chain stretching depends \textit{i)} on the model parameters for the homogeneous chain taken so far (listed in Section 3) and \textit{ii)} on the  cutoff $\phi_{M}$ which regulates the amplitude of the bending fluctuations between adjacent dimers. These issues are investigated by considering a $20$ base pairs heterogeneous fragment whose single strand sequence is: $'\, AAGAAAGGGGGAAAAAAGAA \,'$. With respect to the previous case, $13 \,AT$ base pairs are introduced and this is expected to confer enhanced flexibility to the chain \cite{liang14}. Accordingly the set of potential parameters chosen in the computations is:

$D_{GC}=\,60 meV$, $D_{AT}=\,40 meV$,   $b_{GC}= 3 \AA^{-1}$, $b_{AT}= 2 \AA^{-1}$,  $f_s=\,0.1$, $l_s=\,0.5 \AA$, $K_S=\,10 mev \AA^{-2}$, 
$\rho_{GC, GC} =\,1$, $\alpha_{GC, GC} =\, 2 \AA^{-1}\,$,  $\rho_{AT, GC} =\,1.5$, $\alpha_{AT, GC} =\, 0.2 \AA^{-1}\,$, $\rho_{AT, AT} =\,2$, $\alpha_{AT, AT} =\, 0.1 \AA^{-1}\,$

Essentially, the $AT$ base pairs are modeled by a lower dissociation energy and a larger nonlinear stacking force than the $GC$ base pairs \cite{zdrav06}.

\begin{figure}
\includegraphics[height=8.0cm,width=8.0cm,angle=-90]{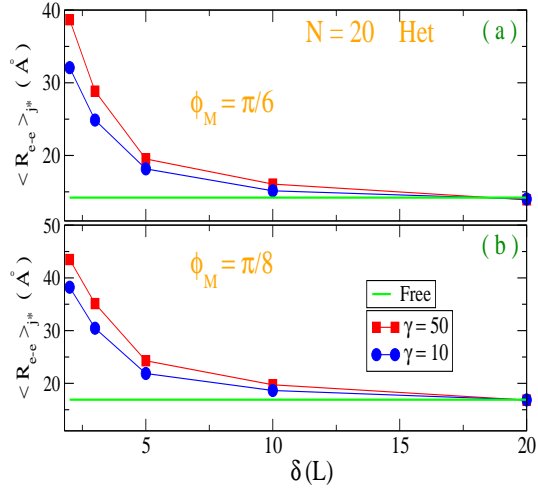}
\caption{\label{fig:3}(Color online) Equilibrium average end-to-end distance for a heterogeneous chain ($20$ base pairs) in a confining channel whose parameters are as in Fig.~\ref{fig:2}. Two values for the bending cutoff, defined in Eq.~(\ref{eq:05}), are assumed in (a)  and (b).   The free molecule end-to-end distances are reported for comparison. 
}
\end{figure}

In Fig.~\ref{fig:3}(a) the average end-to-end distance for the heterogeneous chain is plotted versus $\delta(L)$ setting the same bending cutoff as in Fig.~\ref{fig:2}.
As a main effect, the scale of $< R_{e-e} >_{j*}$ is reduced with respect to Fig.~\ref{fig:2}(b) and, also for the narrowest channel, the heterogeneous molecule is less stretched than its homogeneous kin. 

$< R_{e-e} >_{j*}$ for the unconfined heterogeneous molecule is also reported (green line). This value ($14$ \AA) is smaller than the one previosly found for the homogeneous chain ($17.6$ \AA) pointing to the fact that entropic effects are larger for $AT$ base pairs and cause a more coiled conformation for the helix. 

Reducing the amplitude of the bending fluctuations (Fig.~\ref{fig:3}(b)) amounts to assume a straighter conformation for the free molecule. As a consequence, the channel effectively confines the molecule which appears more elongated than in Fig.~\ref{fig:3}(a) for the same values of $\delta(L)$ and $\gamma$.

In  Figs.~\ref{fig:2},~\ref{fig:3} it has been noted that the $< R_{e-e} >_{j*}$ values for the free and confined DNA molecules converge once the channel width parameter is $\delta(L)=\,20 \AA$ which accordingly has been taken as the upper bound of the x-axis. This convergence physically means that the helical molecule in the pore does not experience any confining effect if its base pair relative distances can fluctuate for lenghts twice as large as the bare helix diameter $R_0$. In this regime, the molecule in the channel behaves as if it were free. 

\begin{figure}
\includegraphics[height=7.0cm,width=8.0cm,angle=-90]{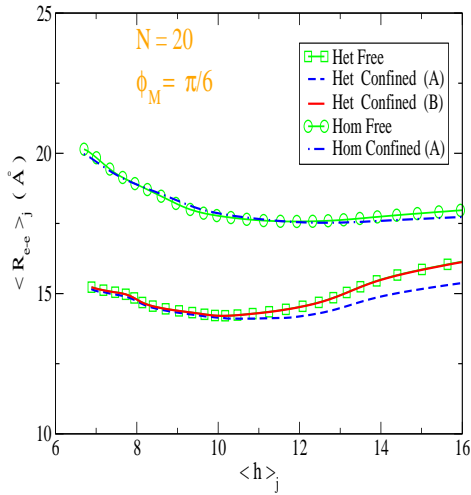}
\caption{\label{fig:4}(Color online) Average end-to-end distance (Eq.~(\ref{eq:09})) versus average helical repeat (Eq.~(\ref{eq:08})). The plots for the free (unconfined) homogeneous and heterogeneous fragments are compared to the plots for the confined molecules. Two sets of channel potential parameters are used. (A) :\, $\delta(L)=\,20 \AA$ and $\gamma=\,10$; (B) :\, $\delta(L)=\,30 \AA$ and $\gamma=\,5$.
}
\end{figure}

Next we investigate whether the convergence between free and confined chains is limited to the equilibrium ($j^*$) conformation of interest or whether it does persist throughout the whole range of helical conformations sampled by our computation. To this purpose we plot in Fig.~\ref{fig:4} the ensemble averaged end-to-end distance in Eq.~(\ref{eq:09}) as a function of the average helical repeat in Eq.~(\ref{eq:08}) both for the homogeneous and heterogeneous chains. The plots for the free chains are compared to the plots for the chains under confinement.  The above discussed equilibrium average end-to-end distances correspond to the minima for $< R_{e-e} >_{j}$.

First we take a set of channel parameters as in Figs.~\ref{fig:2},~\ref{fig:3}, that is the set (A) : \, $\delta(L)=\,20 \AA$ and $\gamma=\,10$.
We observe that the convergence is good for both chains around the minima used in our previous analysis and also in the over-twisting range ($< h >_{j^{}} \, < \, < h >_{j^{*}}$). Instead, for the heterogeneous chain, some discrepancies between free and confined chains are found for the untwisted conformations ($< h >_{j^{}} \,>\, < h >_{j^{*}}$). This stems from the fact that large amplitude fluctuations capable to locally unstack the helix have, for $AT$ base pairs, a larger statistical weight in Eq.~(\ref{eq:05}) (i.e., smaller contribution to the action). Precisely such fluctuations, whose effect is enhanced in the untwisting regime, are  mostly truncated by the confining potential as seen in Fig.~\ref{fig:4}.

The issue is solved by tweaking the channel parameters and setting for example, (B) : \, $\delta(L)=\,30 \AA$ and $\gamma=\,5$. For this choice, the plot for the confined molecule accurately fits the free molecule plot for all possible twist conformations.
Summing up, if the heterogeneous helix, \textit{before entering the pore}, had to be untwisted (e.g., because of thermal effects) with respect to the equilibrium helical repeat, one should have to adjust the potential parameters selecting both hard-wall interaction strength $\gamma$ and  upper bound for $\delta(L)$ which yields convergence between free and confined models. Then, by reducing $\delta(L)$, one could correctly weigh the confinement effect on the molecule size.
This points to the general importance of defining the specific helical conformation in quantitative mesoscopic models for short DNA chains and this is all the more true in order to estimate the real effect brought about by confining potentials.
For the customary case of molecules in the $j^*-$ equilibrium  conformation,  one may adopt the general criterion to set $\gamma$ as the highest interaction strength for which the confining potential fits the free molecule end-to-end distance at a  sufficiently large $\delta(L)$ to allow for broad base pair fluctuations. 

Finally it is worth pointing out that, in Fig.~\ref{fig:4}, the minima $< R_{e-e} >_{j*}$  occur for $< h >_{j^{*}}$'s close to (albeit not coincident with) the usual value measured for long DNA chains. This is an interesting output of the computational method, given the broadness of the range of possible helical conformations initially sampled by the code. Although the helical pitch of a molecule may in general vary according to sequence specificities and bending conformation, the expected value for short chains should be, under physiological conditions, in the range of the above mentioned $h^{exp}$ \cite{shore83}.

\section{ Conclusions}

We have addressed the properties of short DNA molecules in a cylindrical pore with the purpose of building a potential which weighs the confinement effects on the base pair fluctuations. Such effects have been analyzed through a 3D mesoscopic model appropriate to describe the DNA flexibility properties and, specifically, to account for the global molecule size as measured by the end-to-end distance.  It is found that the average end-to-end distance for a confined homogeneous chain markedly grows, by narrowing the cylinder diameter, over the value computed for the free chain. Consistently, the molecule elongation is accompanied by the expected entropic reduction peculiar of the more ordered conformation. The stretching occurs, albeit reduced, also for a  sequence rich in AT base pairs which confer enhanced chain flexibility and induce a coiled helical conformation. It is inferred that the optimal pore diameter required to stretch heterogeneous molecules should decrease in the presence of a high percentage of AT base pairs.
We have also shown how the quantitative results for the end-to-end distance may depend on the model parameters, markedly on the strength of the confining potential which has been taken as a tunable parameter. In this regard, measurements of the stretching of short molecules in pores, currently not available, could contribute to test our predictions and set the model parameters for specific sequences. 
Finally, as a distinctive feature of our method, it is noticed that the computed helical pitches for the entropically favored twist conformations, $< h >_{j^{*}}$ in Fig.~\ref{fig:4},  are close to the average experimental data usually found for long DNA chains at room temperature. Thus, the whole of the obtained results indicate that our Ansatz for the hard-wall potential can suitably model a confining channel which exerts a repulsive force on the helical chain, shrinks the amplitude of the base pair fluctuations and straightens the molecule.


\begin{thebibliography}{widest-label}

\bibitem{seeman17}
N.C. Seeman, H.F. Sleiman, \textit{ Nat. Rev. Mater.} \textbf{3}, (2017) 17068.


\bibitem{yang16}
 Y. Yang, J. Wang, H. Shigematsu, W. Xu, W.M. Shih, J.E. Rothman, C. Lin,  \textit{ Nat. Chem.} \textbf{8},  (2016) 476-483.


\bibitem{io18b}
M. Zoli, \textit{EPL} {\bf 123},  (2018) 68003.



\bibitem{busta06} 
J. Gore, Z. Bryant, M. N\"{o}llmann, M.U. Le, N.R. Cozzarelli and C. Bustamante, \textit{ Nature}  \textbf{442},  (2006) 836-839.

\bibitem{biton18}
Y.Y. Biton, \textit{J. Chem. Theory Comput.}  \textbf{14},  (2018) 2063-2075.


\bibitem{xia08}
Q. Xia, K.J. Morton, R.H. Austin, S.Y. Chou, \textit{Nano Lett.} \textbf{8},  (2008) 3830-3833.


\bibitem{gennes77}
M. Daoud, P. G. de Gennes, \textit{ J. Phys.} \textbf{38},  (1977) 85-93.

\bibitem{odijk83}
T. Odijk, \textit{Macromolecules}  \textbf{16},  (1983) 1340-1344.


\bibitem{chen13} 
Y. Chen, K. Luo, \textit{J. Chem. Phys.} \textbf{138},  (2013) 204903.

\bibitem{pablo07}
T.A. Knotts, N. Rathore, D.C. Schwartz, J.J. de Pablo, \textit{J. Chem. Phys. } {\bf 126},  (2007) 084901.



\bibitem{io11}
M. Zoli,       \emph{J. Chem. Phys.}  \textbf{135},   (2011) 115101.


\bibitem{coll95}
F. Zhang, M.A. Collins,    \emph{Phys. Rev. E} {\bf 52},  (1995) 4217-4224.

\bibitem{albu14}
D.X. Macedo, I. Guedes, E.L. Albuquerque, \textit{Physica A} \textbf{404},  (2014) 234-241.




\bibitem{pey93}
T. Dauxois, M. Peyrard, A.R. Bishop,  \emph{Phys. Rev. E}  \textbf{47},   (1993) R44-47.

\bibitem{io12}
M. Zoli, \textit{J. Phys.: Condens. Matter} {\bf 24},    (2012) 195103.


\bibitem{zhang97}
Y.L. Zhang, W.M. Zheng, J.X. Liu,  Y.Z. Chen,   \textit{ Phys. Rev. E}  \textbf{56},  (1997) 7100-7115.


\bibitem{kalos09}
G. Kalosakas, S. Ares,     \textit{J. Chem. Phys.}   {\bf 130},  (2009) 235104.

\bibitem{lucia98}
J. SantaLucia Jr.,  \textit{Proc. Natl. Acad. Sci. USA} \textbf{95},  (1998) 1460.

\bibitem{campa98}
A. Campa, A. Giansanti, \textit{Phys. Rev. E} \textbf{58},  (1998) 3585.


\bibitem{krueg06}
A. Krueger, E. Protozanova, M.D. Frank-Kamenetskii,  \textit{Biophys. J.}  {\bf 90},  (2006) 3091-3099.


\bibitem{io18c}
M. Zoli,       \emph{J. Chem. Phys.}  \textbf{148},  (2018) 214902.


\bibitem{io13}
M. Zoli,       \emph{J. Chem. Phys.}  \textbf{138},  (2013) 205103.

\bibitem{weber15}
I. Ferreira, T.D. Amarante,  G. Weber,    \emph{J. Chem. Phys.} \textbf{143}, (2015) 175101.

\bibitem{io14b}
M. Zoli,  \textit{J. Chem. Phys.} {\bf 141},  (2014) 174112.

\bibitem{io16b}
M. Zoli,   \textit{J. Chem. Phys.}   {\bf 144},   (2016) 214104. 


\bibitem{fehi}
R.P. Feynman,  A.R. Hibbs,   {\it Quantum Mechanics and Path Integrals}, (Mc Graw-Hill, New York,  1965).

\bibitem{singh11}
S. Srivastava, N. Singh,  \emph{J. Chem. Phys.} \textbf{134},   (2011) 115102.

\bibitem{kim14}
T.T. Le, H.D. Kim,    \emph{Nucleic Acids Res.}   \textbf{42},  ( 2014) 10786-10794.

\bibitem{vafa}
R. Vafabakhsh, T. Ha,  \emph{Science}  \textbf{337},    (2012) 1097-1101.

 
\bibitem{io19}
M. Zoli, \textit{Phys. Chem. Chem. Phys.}  {\bf 21},   (2019) 12566-12575.

\bibitem{wang79}
J.C. Wang,   \emph{Proc. Natl. Acad. Sci. USA}   \textbf{76},  (1979) 200-203.

\bibitem{io17}
M. Zoli, \textit{J. Phys.: Condens. Matter} {\bf 29},   (2017)  225101.

\bibitem{liang14}
H. Li, Z. Wang,  N. Li,  X. He,  H. Liang,  \textit{J. Chem. Phys. } {\bf 141},  (2014) 044911.

\bibitem{zdrav06}
S. Zdravkovi\'{c},  M.V. Satari\'{c},  \emph{Phys. Rev. E}  {\bf 73},   (2006) 021905.


\bibitem{shore83}
D. Shore, R.L. Baldwin,   \emph{J. Mol. Biol.} \textbf{170},    (1983) 957-981; ibid.,   \textbf{170},  (1983) 983-1007.














\end{thebibliography}
\end{document}